\documentclass[aps,preprint,showpacs,superscriptaddress]{revtex4}
                                                                                
\usepackage{graphicx} 
                                                                                
\begin{document}

\title
{Half metallic digital ferromagnetic heterostructure 
composed of a $\delta$-doped layer of Mn in Si}
                                                                                
\author{M. C. Qian}
\author{C. Y. Fong}
\author{Kai Liu}
\author{Warren E. Pickett}      
\affiliation{Department of Physics, 
University of California, Davis CA 95616-8677}
\author{J. E. Pask}
\author{L. H. Yang}
\affiliation{Lawrence Livermore National Laboratory, University of California, Livermore, CA 94551}

\date{\today}
                                                                                
\begin{abstract}
We propose and investigate the properties of 
a digital ferromagnetic heterostructure (DFH)
consisting of a $\delta$-doped layer of Mn in Si, 
using \textit{ab initio} electronic-structure methods.
We find that (i) ferromagnetic order of the Mn layer 
is energetically favorable 
relative to antiferromagnetic,
and (ii) the heterostructure is a two-dimensional half metallic system.
The metallic behavior is contributed by three majority-spin bands
originating from hybridized Mn-$d$ and nearest-neighbor Si-$p$ states,
and the corresponding carriers 
are responsible for the ferromagnetic order in the Mn layer. 
The minority-spin channel has a calculated semiconducting gap of 0.25 eV.
Analysis of the total and partial densities of states, 
band structure, Fermi surfaces and associated charge density 
reveals the marked two-dimensional nature of 
the half metallicity. 
The band lineup is found to be favorable for retaining the half metal
character to near the Curie temperature ($T_{C}$).
Being Si based and possibly having a high $T_{C}$
as suggested by an experiment on dilutely doped Mn in Si, 
the heterostructure may be of special interest 
for integration into mature Si technologies for 
spintronic applications.
\end{abstract}
                                                                                
\pacs{75.50.Pp, 71.15.Mb, 71.70.Gm, 72.80.Cw}

\maketitle

Doping magnetic transition metal elements 
into semiconductors has attracted much 
attention recently, especially the doping of Mn in GaAs\cite{ohno}, 
which has inspired 
much interest in the new and immensely promising field of spintronics
\cite{spin1,spin2,spin3}. Since 2000, a new class of potential spintronic materials 
has been discovered: the half metallic transition metal pnictides 
with zincblende structure. Two 
such compounds, CrAs\cite{akinaga} and CrSb\cite{zhao}, 
have been grown in thin film form. 
Many other pnictides and carbides\cite{pask,galanakis,liu}
have been predicted based on first-principles calculations in the generalized 
gradient approximation (GGA)\cite{PBE}. 
Related quantum structures, such as superlattices\cite{fong1,fong2,fong3}
and quantum dots\cite{fong4}, have also 
been studied. Sanvito and Hill\cite{ssan} 
investigated the half metallic properties of a digital 
ferromagnetic heterostructure (DFH) 
composed of a layer of Mn in GaAs. 
The key property of all such half metallic materials is 
the 100 \% spin polarization at the Fermi energy, $E_{F}$. 
Therefore, the magnetoresistance (MR) effect is expected to
be much larger and becomes infinite in the ideal case.
Devices made of these materials, such as MR sensors, memories and
switches, will have superior qualities to the ones presently available.
However, the realization of devices based on these materials 
has been hindered by difficulties in growth and fabrication processes involving 
III-V compounds. 
Fabrication techniques for Si based devices are more mature; and so they 
might be more readily manufactured, if corresponding half metallic devices
could be engineered. 
The prediction of 
ferromagnetic order 
in Mn-doped bulk Si\cite{nanjing} and 
the growth of dilutely doped Mn in Si\cite{bolduc, du}
have been reported.
The experimental results are particularly encouraging 
because the measured Curie temperature, $T_C$, 
is over 400 K, in stark contrast to Mn-doped GaAs 
which suffers from a $T_C$ far 
below room temperature. 
A critical shortcoming of Mn-doped GaAs with respect to spintronic applications 
might thus be surmounted by such Si based materials.

Here, we report our design of a DFH consisting of 
a $\delta$-doped layer of Mn in a Si 
substrate (Mn/Si-DFH), with a $\delta$ function (single layer)
doping profile along the growth direction, and our prediction that this DFH is 
a two-dimensional half metal. We find that the metallic properties are 
contributed by extended states
originating from hybridized Mn-$d$ and nearest-neighbor 
Si-$p$ majority-spin states,
while the minority-spin channel is semiconducting with a 0.25 eV gap.

We employed planewave pseudopotential density-functional methods\cite{kresse}, 
in the generalized gradient approximation (GGA) 
to exchange and correlation\cite{PBE}. 
Ultra-soft pseudopotentials\cite{van} were used 
to facilitate the accurate 
treatment of transition metal atoms.
The heterostructure was modeled by a unit cell consisting of 32 atoms  
in layers along the [001]-direction of epitaxial growth,
as shown in Fig.~1.  
The unit cell is tetragonal with $\mathbf{a}$ and $\mathbf{b}$ axes along the 
[110]- and [\={1}10]-directions, and lattice constant $a_{0}/\sqrt{2}$, where 
$a_{0}=5.45$ \AA ~~is the optimized lattice constant of Si crystal.
The indices are defined with respect to the conventional cell.
We have checked the effect of the thickness of the
Si region from 15 to 39 Si layers along the [001]-direction.
The half metallic properties and the density states of
the majority-spin at $E_{F}$ remain unaffected.
The gap between the lowest unoccupied state
in the minority-spin channel and $E_{F}$ is 0.15 eV
at 15 Si layers separation
and stabilized to 0.16 eV 
at separations of 31 Si
layers or more. Further details will be presented elsewhere\cite{qian}.
The separation between Mn layers in neighboring cells was sufficient to make 
cell-cell interactions negligible, as we discuss below.
A planewave cutoff of 450 eV and 
Monkhorst-Pack\cite{monk} mesh of $11 \times 11 
\times 1$ {\bf k}-points was used in all calculations.
Larger {\bf k}-point sets were employed to verify convergence of the above 
set to the order of 1 meV/atom.
Wigner-Seitz sphere radii of 1.35~\AA~ and 1.32~\AA~ were used for Mn and 
Si atoms, respectively, to construct projected densities of states.
Atomic positions were optimized by conjugate-gradient minimization of 
the total energy.

The magnetic moment per unit cell, density of states 
at the Fermi energy, and 
energy difference between ferromagnetic and antiferromagnetic ordering
for the unrelaxed and relaxed cases are given in Table~1. 
Without lattice relaxation, 
the magnetic moment is an integer, 3.0 $\mu_{B}$. 
There is a gap of 0.20 eV in the minority-spin states. 
The effect of lattice relaxation is small 
because the optimized lattice constant of Si (5.45 \AA) 
is close to that of (hypothetical) 
zincblende MnSi (5.48 \AA)\cite{qian}.  
The gap for the relaxed case opens slightly to 0.25 eV. 
Both with and without relaxation, 
the ferromagnetic phase has lower energy 
than the antiferromagnetic phase.
Since the relaxation effect will not alter our conclusions about DFH,
in the following, we focus our analysis on the unrelaxed case. 

\begin{table}
\caption{Comparison of the unrelaxed and relaxed cases for Mn 
$\delta$-doped in Si.
The magnetic moment $m$, density of states at the Fermi energy $N(E_{F})$ 
per eV-unit cell, ferromagnetic--antiferromagnetic energy difference $E_{FA}$, 
and minority-state gap $E_g$ are listed. The difference $E_{FA}=E_{FM}-E_{AFM}$ 
between FM and AFM ordering is calculated for the unit cell 
with one Mn-Mn pair.
For both cases, the FM ordering has lower energy.}
\begin{ruledtabular}
\begin{tabular}{ccccc}
      &$m(\mu_{B})$&$N(E_{F})$&$E_{FA}$(meV)&$E_{g}$(eV)  \\ \hline
Unrelaxed&3.0&1.06 &-523.91 &0.20\\  \hline
Relaxed  &3.0&1.25 &-442.38 &0.25   \\ 
\end{tabular}
\end{ruledtabular}
\end{table}

The total and partial densities of states (PDOS) 
for the majority- and minority-spin channels are shown in Fig.~2. 
The PDOSs show projections onto the Mn, 
nearest Si (Si(I)), and farthest Si (Si(II)) atoms.
Considering first the majority-spin states, we find
a finite density of states 
at $E_{F}$, as shown in the top panel. 
The states in a 3 eV range around $E_{F}$ 
(from $\sim -1$ to $+2$ eV) show 
strong hybridization between Mn-t$_{2g}$ and 
Si(I)-$p$ states (second and third panels).
In contrast, there is no significant contribution 
from the more distant Si(II) atoms.
The bonding states centered $\sim 2.6$ eV below $E_{F}$ 
exhibit correspondingly strong 
Mn-t$_{2g}$--Si(I)-$p$ character. 
The nonbonding Mn-e$_{g}$ states are located $\sim 2.5$ eV below $E_{F}$, 
concentrated in a somewhat narrower range of about 1 eV. 
The conduction states $\sim 2$ eV above $E_{F}$ and higher are contributed 
mainly by the Si atoms. 
Considering now the minority-spin channel, we find
a semiconducting gap of $\sim 0.20$ eV between valence and conduction bands. 
The valence states are mainly bulk Si, with one band 
of Mn-t$_{2g}$--Si(I)-$p$ character 
at $\sim 1$ eV below $E_{F}$. 
The conduction states originate 
from both antibonding $p$--t$_{2g}$ hybrid states and nonbonding e$_{g}$ states 
of the Mn, leading to a peak at $\sim 0.5 $ eV.
The more distant Si(II) atoms contribute a broad manifold 
extending to 4 eV below $E_{F}$, 
much as in Si crystal. 
The results indicate that
the interaction of the Mn atom is confined 
primarily to its nearest-neighbor Si(I) atoms.

The band structure along symmetry lines
in the {\bf k}$_{x}$-{\bf k}$_{y}$ plane ($\Gamma$-R-X-$\Gamma$) 
and along the {\bf k}$_{z}$-direction ($\Gamma$-Z) 
in the energy range $-1.5$ to 1.5 eV is shown in Fig.~3. 
The sizes of the circles
indicate the contribution from Mn-$d$ states: larger circles indicate 
larger contributions. 
In Fig.~3(a), three majority-spin bands pass through the Fermi energy 
in the {\bf k}$_{x}$-{\bf k}$_{y}$ plane.
We label these bands 1, 2 and 3.
The lowest energy states of the three bands are at the R point 
and are occupied.
Around the $\Gamma$ point, the three bands are unoccupied. 
As discussed above in PDOS, 
the states in the vicinity of $E_{F}$ contributing to the conduction
originate from the hybridization of Mn-t$_{2g}$ 
and (near-neighbor) Si(I)-$p$ states. 
Along $\Gamma$-Z, the marked flatness of the associated bands indicates the 
negligible interaction between Mn planes: the states are extended along the 
planes and well localized perpendicular to them,
forming a metallic two dimensional sheet of majority carriers only.
Compared to the corresponding bands along $\Gamma$-X of 
the Mn/GaAs DFH studied by Sanvito and Hill \cite{ssan},
the partially filled bands in Mn/Si-DFH are substantially broader, 
reflecting the more ionic character of the As atoms relative to Si.
In Fig.~3(b),
the band structure for the minority-spin channel is shown. 
The semiconducting gap is indirect with the top of the
valence band near the $\Gamma$ point and the bottom of the 
conduction band at the zone corner R point.
From the above, then, it is clear that the carriers in the Mn/Si-DFH 
come from three majority-spin bands,
and mediate the exchange interactions between the local magnetic moments
which give rise to the ferromagnetic order in the Mn layer.
The band structures given in Fig.~3 (a) and (b) exhibit 
clearly the marked two-dimensional half metallic character
of the heterostructure.

In order to better characterize the relevant conduction states,
we examine the Fermi surfaces and associated 
charge density in Fig.~4.
The band 1 
forms a hole pocket, a surface with closed orbit,
around the $\Gamma$ point. 
Bands 2 and 3 form two electron surfaces centered at the R point
in the {\bf k}$_{x}$-{\bf k}$_{y}$ plane. 
In Fig.~4(b), as an example for illustration, 
the corresponding majority-spin hole charge densities are 
obtained by integrating states in the vicinity of $E_{F}$ up to 0.5 eV. 
It can be seen that the hole states are strongly confined 
in the vicinity of the Mn layer and distributed in the
bonding directions between Mn and Si(I) atoms.

One important question is whether the half metallic character can be 
maintained up to $T_{C}$ in this DFH.
Based on an analysis of available experimental data on NiMnSb alloys,
Hordequin {\it et al.} \cite{hor} identified the temperature $T^*$ (80 K)
well below $T_C$ (730 K)
at which NiMnSb undergoes an electronic phase transition
from half metal to normal ferromagnet.
This crossover occurs because the minority bandgap is the result of 
exchange splitting of bands, which is proportional to the magnetization 
and decreases as temperature increase.
The interpretation is that one of the gap edges crosses the Fermi level
as the gap decreases, thereby losing the half metal character.
From Hall measurements, they found that the hole concentration increases
with temperature and that the low
$T^*$ is the signature of 
spin flip transitions from the majority-spin states at $E_{F}$ 
to the bottom of the conduction band of the minority-spin channel,
which become allowed as the gap decreases. 
They concluded that $T^{*}$ could reach $T_{C}$
if $E_{F}$ lies near the top of the valence band of the minority-spin channel,
rather than near the bottom of the conduction band. 
Unlike NiMnSb, 
this advantageous band lineup does occur 
in the band structure of Mn/Si-DFH shown in Fig.~3. 
Fig.~3(c) shows a schematic diagram of density of states of Mn/Si-DFH 
near the $E_{F}$. 
The spin-flip gap $\delta$ between the Fermi energy and 
the bottom of the conduction band in the minority-spin channel is about 0.16 eV,
which is close to the semiconducting gap $\Delta=0.2$ eV 
from the valence band maximum (VBM) to the bottom of the conduction band
in the minority-spin channel.
In addition, the bottom of the conduction band is well below
the conduction band minimum (CBM) of host Si crystal. 
At finite temperature, there are two possible spin-flip 
transitions due to the interaction between the spin of a carrier
and the fluctuation of local moments of the Mn atoms. 
One is the transition from the top of valence states 
in the minority-spin channel to the majority states at $E_{F}$,
but this will be strongly suppressed by matrix element effects because
the uppermost valence states are bulk Si bonding states whereas
the states at $E_{F}$ are Mn $d$ - Si(I) $p$ hybridized and they 
only weakly overlap bulk Si states.
Thus, as in NiMnSb, only the second one dominates the spin-flip scattering
and deteriorates the half metallicity,
because both the majority states 
at E$_{F}$ and the minority states at the bottom of conduction band
are coming from the hybridized Mn $d$ - Si $p$ states.
This second process, 
of an electron at the $E_{F}$ undergoing a spin flip
transition to the bottom of conduction band, 
involve the so-called spin-flip gap $\delta$.
It should be mentioned that the GGA usually underestimate the semiconducting
gap of Si crystal, and a wider gap $\delta$ should be more realistic.
Therefore, for the Mn/Si-DFH, 
the half metallic character could persist up to a temperature 
comparable to $T_{C}$. 

In summary, using first-principles electronic-structure methods, 
we have designed 
a low strain DFH composed of a Mn $\delta$-doped layer in Si substrate, 
which  exhibits two-dimensional half metallic properties.
The metallic states are hybrid states originating 
from Mn-t$_{2g}$ and nearest-neighbor Si-$p$ states 
in the majority-spin channel.
A semiconducting gap is retained
between minority-spin bonding and antibonding states. 
From the DOS and band structure, the half metallic properties
are shown to have marked two-dimensional character 
and are likely to persist up to $T^{*} \approx T_C$, 
in marked contrast to NiMnSb.

Since the measured $T_C$ in dilutely doped Mn in bulk Si is 
approximately 400 K\cite{bolduc}, the
$T_C$ for this ordered DFH may be substantially higher than room temperature. 
Due to the maturity of Si technologies, it may be expected that 
Si based devices will be more readily fabricated 
than their GaAs based counterparts. 
If the designed DFH can be grown, such Si based half metallic 
materials could lead to a breakthrough in 
the realization of spintronic devices 
in the near future, and to a new 
generation of devices in the years to come.

This work is partially supported by National Science Foundation 
with grant No. ESC-0225007, and the San Diego Supercomputer Center. 
This work was also performed, in part, under the auspices of 
the U. S. Department of Energy by the University of California, 
Lawrence Livermore National Laboratory 
under contract No. W-7405-Eng-48.

\newpage
\begin{figure}
\includegraphics[scale=0.60,angle=0]{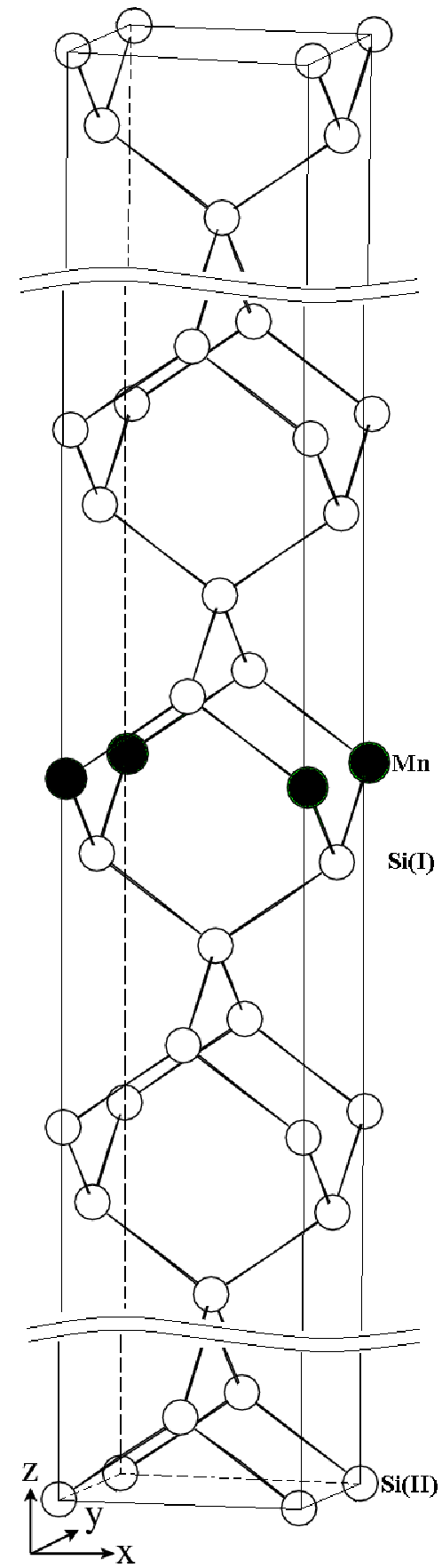}
\caption{Mn/Si-DFH unit cell, consisting of 32 layers in the z-direction. Black circles denote Mn; open circles denote Si.
}\label{fig:f1}
\end{figure}
                                                                                
\newpage
\begin{figure}
\includegraphics[scale=0.70,angle=0]{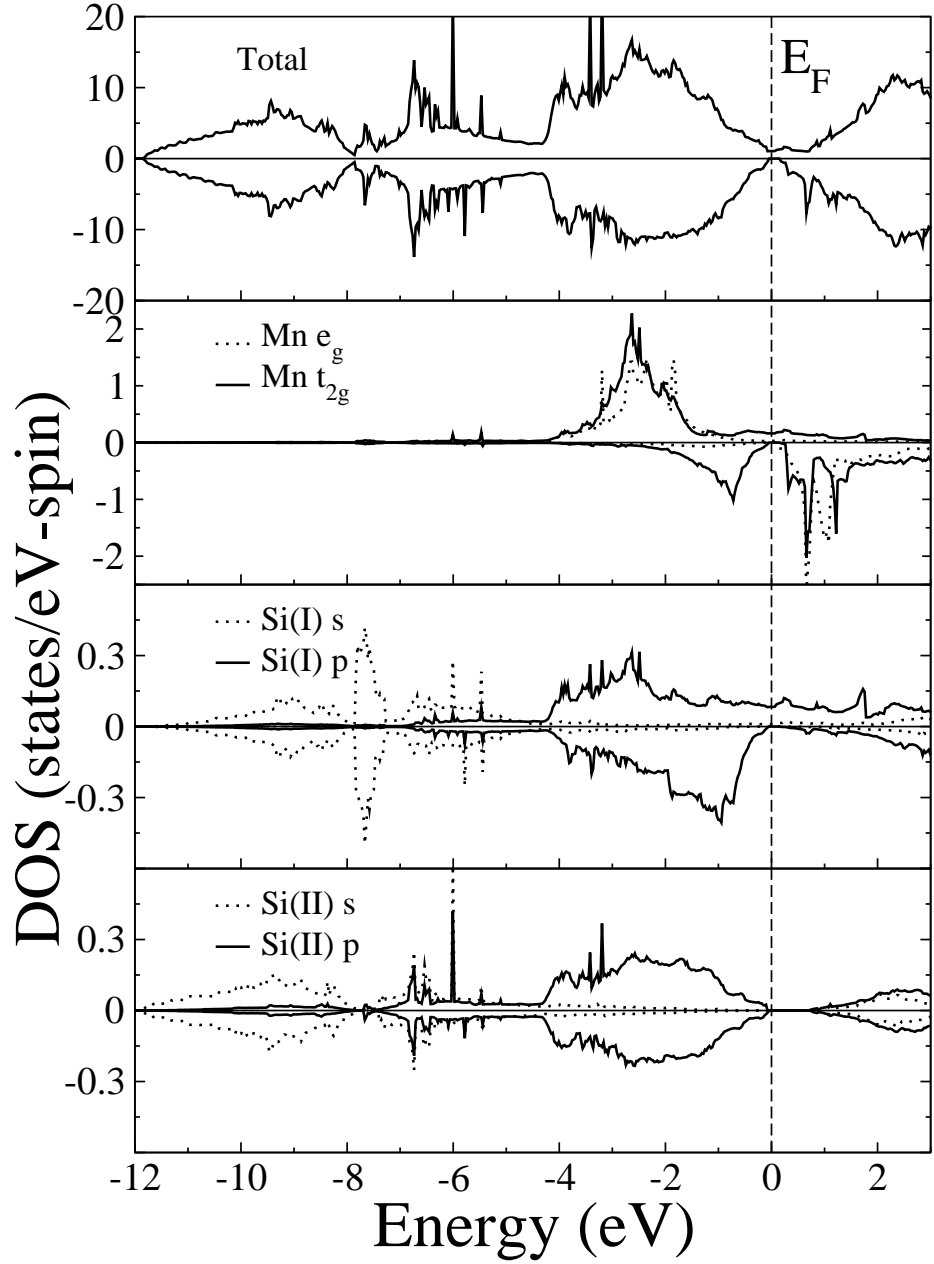}
\caption{Calculated total and partial 
densities of states for Mn $\delta$-doped in Si.
Si(I) and Si(II) refer to nearest and farthest silicons
from Mn. Majority densities are plotted as positive values; minority
densities, as negative.
The vertical dashed line indicates the Fermi energy.}
\end{figure}

\newpage
\begin{figure}
\centerline{
\begin{tabular}{cc}
\multicolumn{1}{c}{
\includegraphics[scale=0.38,angle=270]{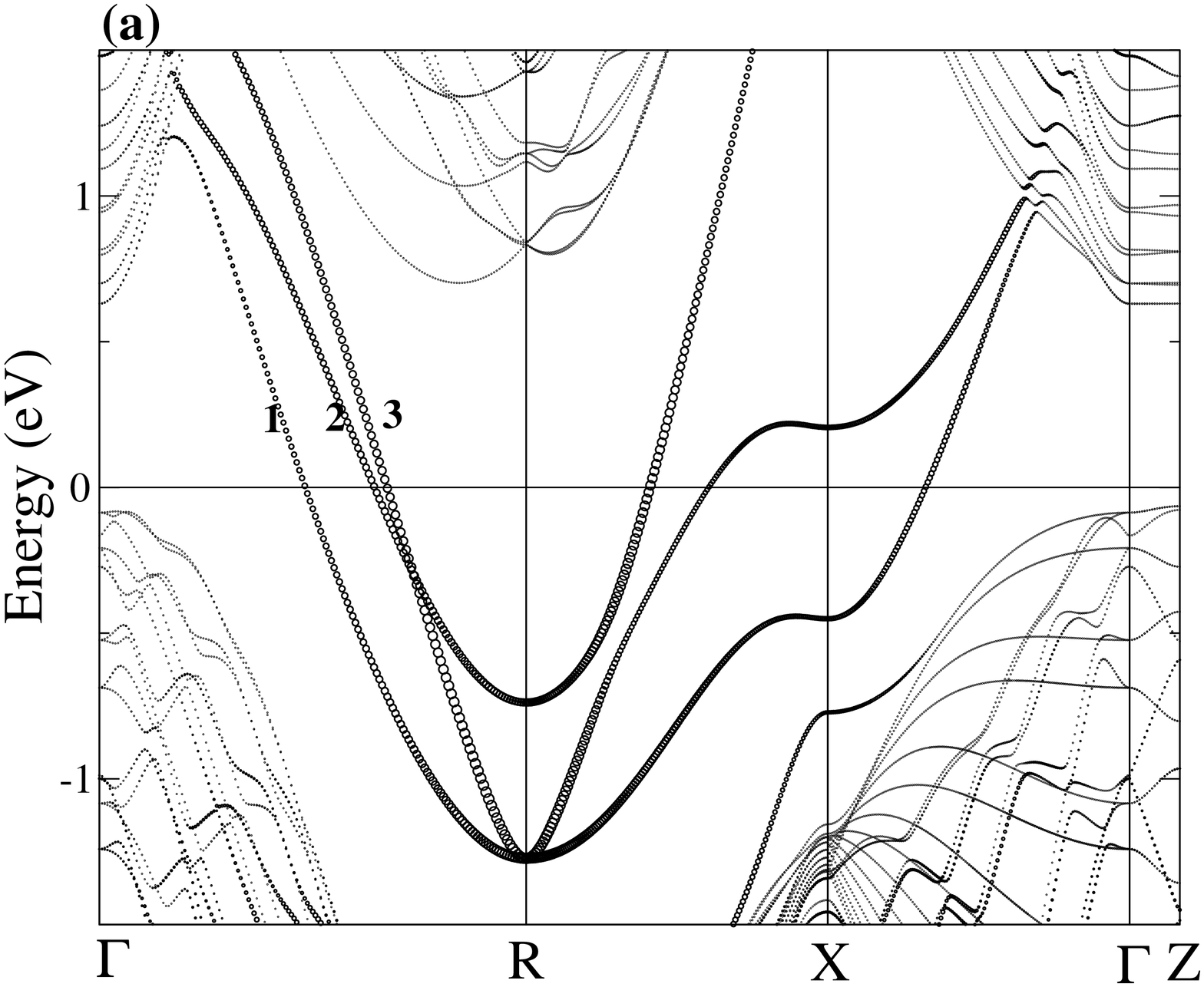}
}
&
\multicolumn{1}{c}{
\includegraphics[scale=0.38,angle=270]{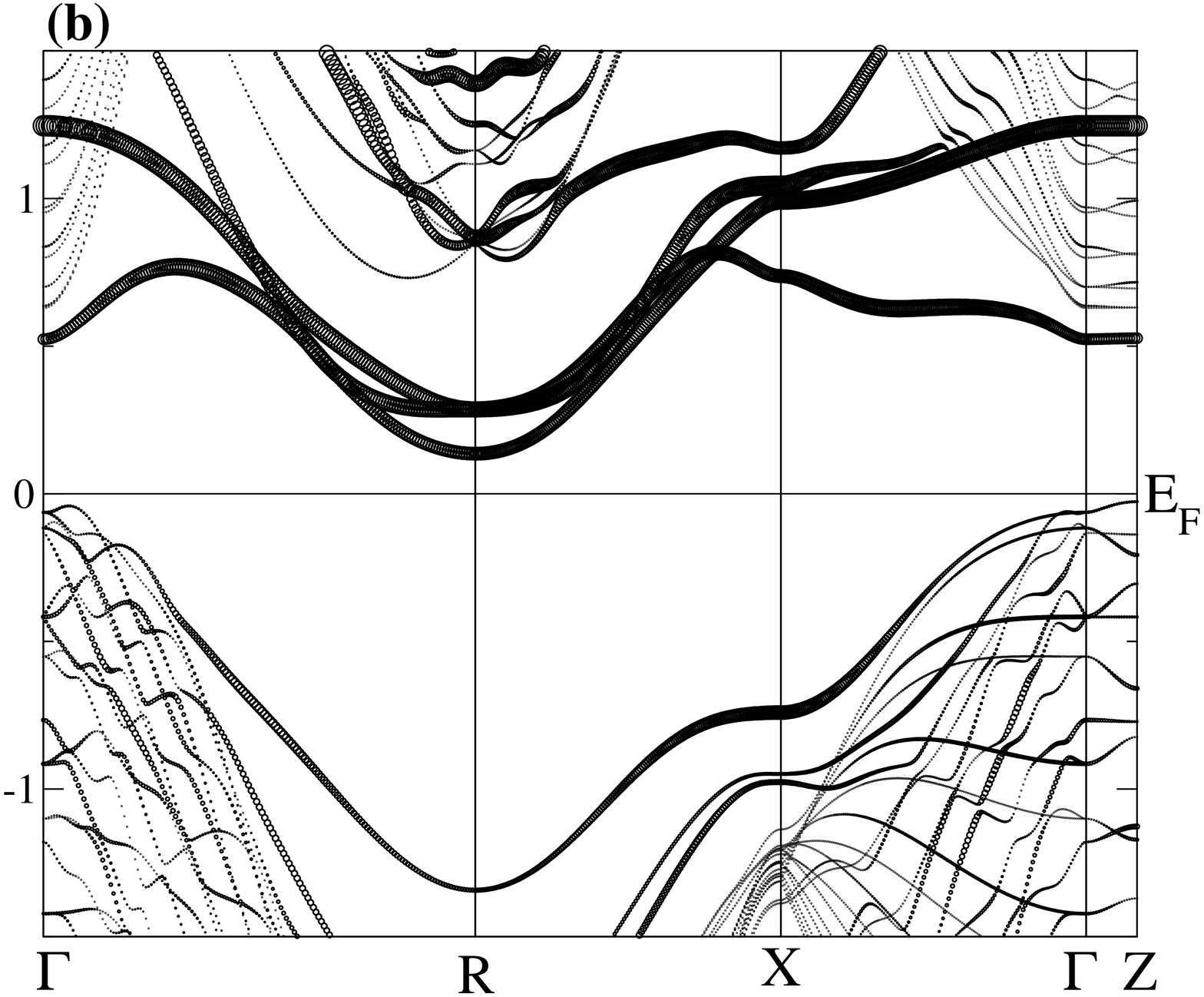}
} \\
\multicolumn{2}{c}{
\includegraphics[scale=0.80,angle=0]{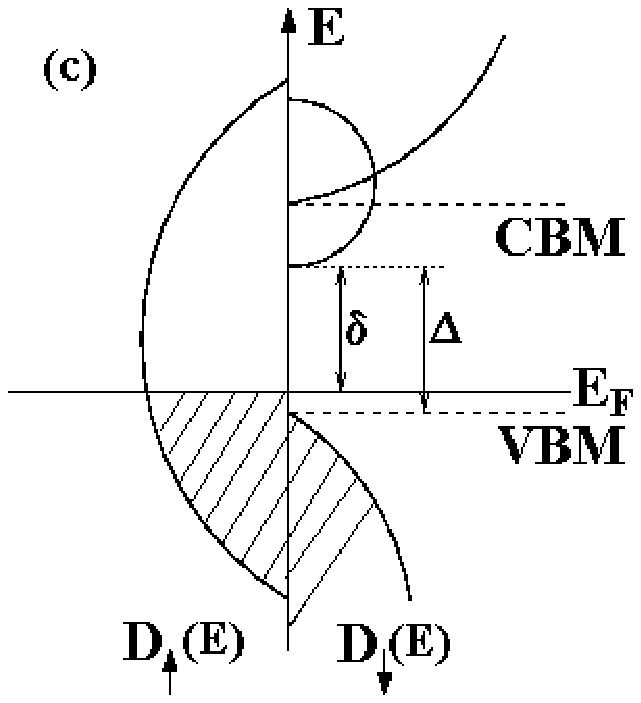}} 
\end{tabular}
}
\caption{Calculated band structure along $\Gamma$-R-X-$\Gamma$ in the {\bf k}$_{x}$-{\bf k}$_{y}$ plane and along $\Gamma$-Z 
in the {\bf k}$_{z}$-direction, (a) spin up and (b) spin down
for Mn $\delta$-doped in Si.
High symmetry points of the Brillouin zone are
$\Gamma=(0,0,0)$, R$=(1,1,0)$, X$=(1,0,0)$, and Z$=(0,0,1)$.
The size of circles indicates the fraction of Mn $d$ character.
The conduction bands are labeled 1, 2 and 3.
(c) Schematic diagram of the density of states D(E) 
in Mn/Si-DFH of the two spin channels near $E_{F}$. 
VBM and CBM are the valence band maximum
and conduction band minimum, respectively, of Si crystal for comparison.} 
\end{figure}

\newpage
\begin{figure}
\centerline{
\begin{tabular}{ll}
{\Large \bf (a) } & {\Large \bf (b) }  \\
\multicolumn{1}{c}{
\includegraphics[scale=0.45,angle=0]{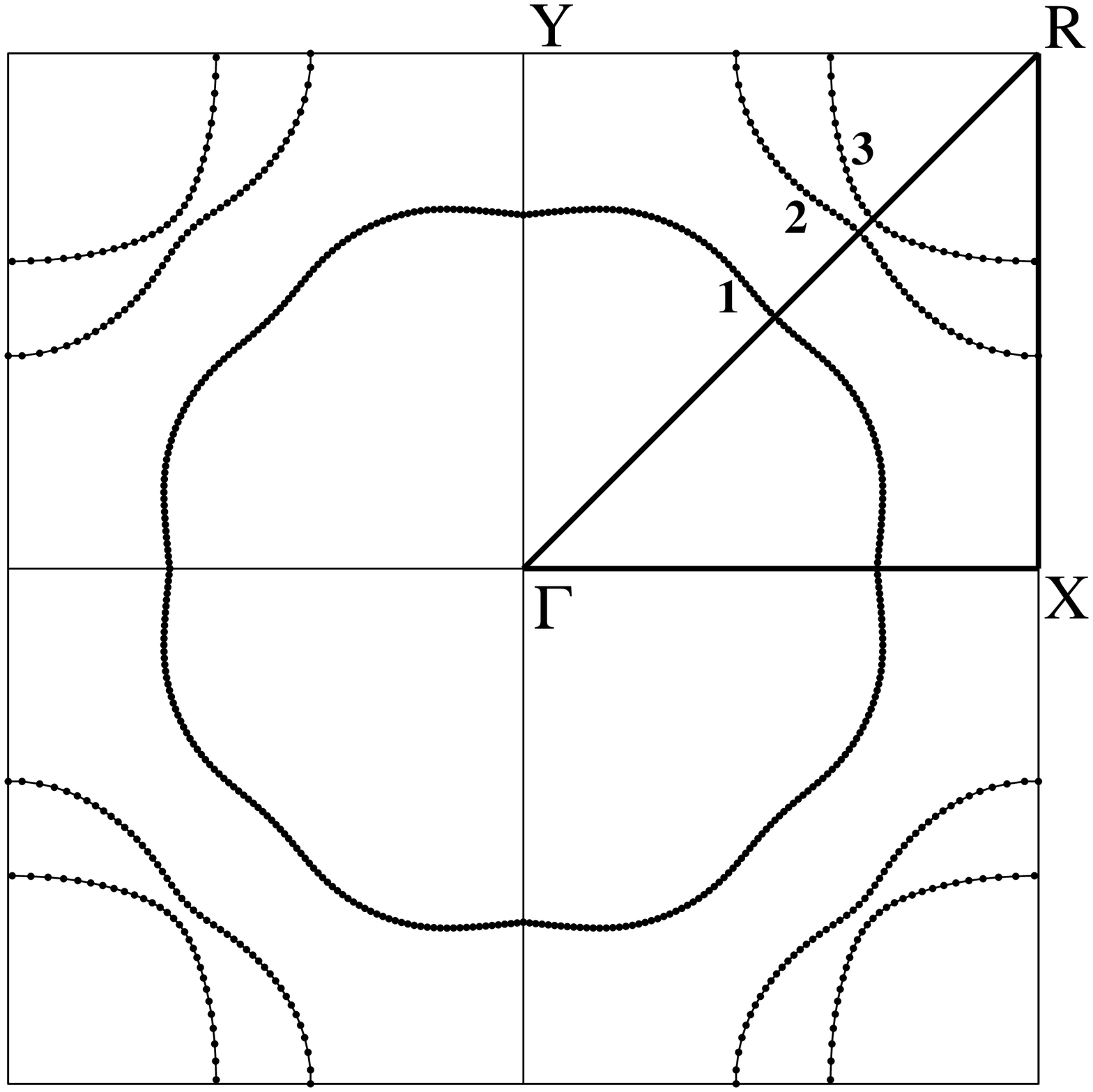}
} & 
\multicolumn{1}{c}{
\includegraphics[scale=0.80,angle=0]{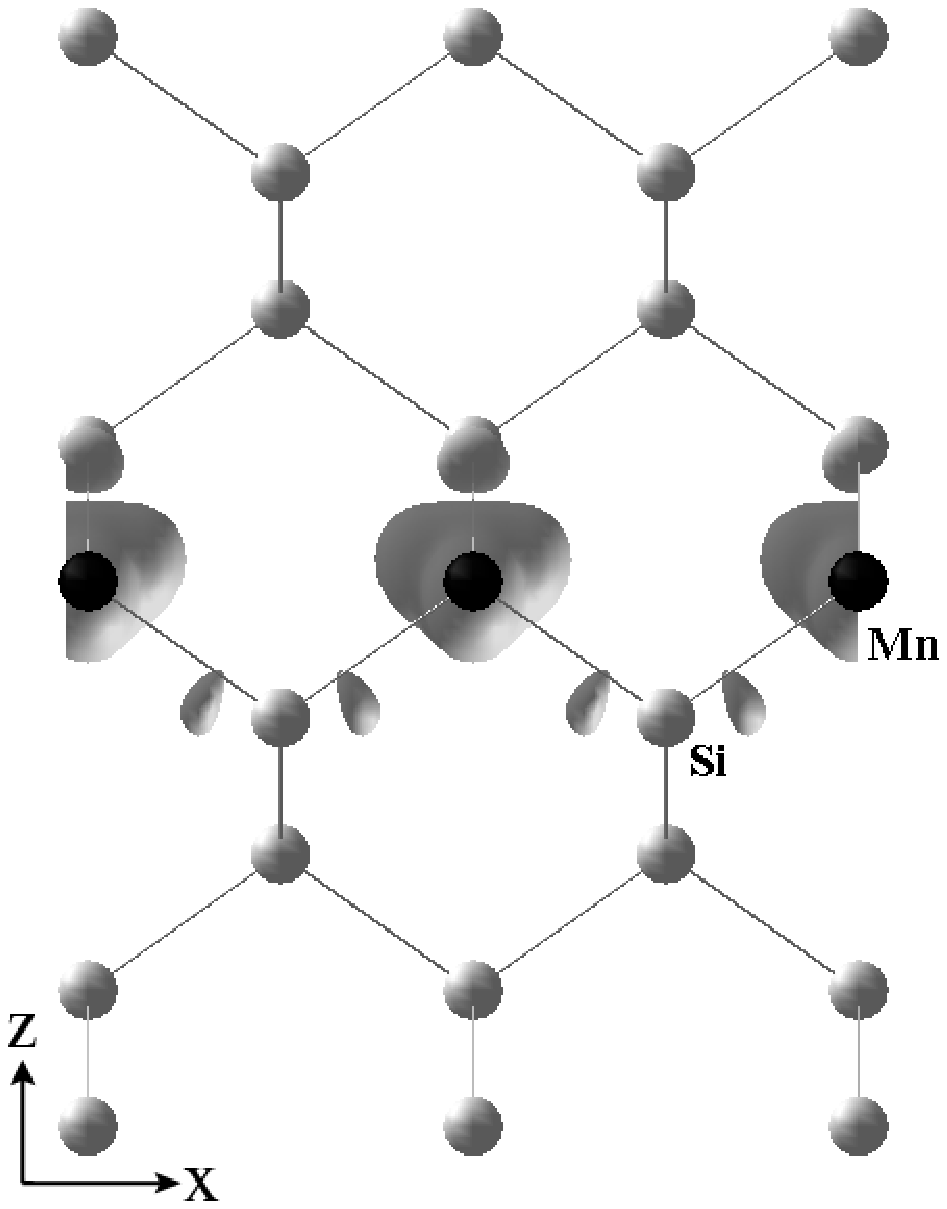}
} \\
\end{tabular}
}
\caption{(a) Calculated two-dimensional Fermi surfaces. 
(b) Majority-spin hole charge density 
in the vicinity of $E_F$ for Mn $\delta$-doped in Si.}
\end{figure}


\begin{thebibliography}{20}

\bibitem{ohno}
Y. Ohno, D. K. Young, B. Beschoten, F. Matshkura, H. Ohno, 
and D. Awschalom, Nature {\bf 402}, 790 (1999).

\bibitem{spin1} G. A. Prinz, Science {\bf 282}, 1660 (1998).

\bibitem{spin2} P. Ball, Nature {\bf 404}, 918 (2000).

\bibitem{spin3} W. E. Pickett and J. S. Moodera, 
Physics Today {\bf 54}, 39 (2001).

\bibitem{akinaga}
H. Akinaga, T. Manago, and M. Shirai, 
Jpn. J. Appl. Phys. {\bf 39}, L1118 (2000).

\bibitem{zhao}
J. H. Zhao, F. Matsukura, T. Takamura, E. Abe, D. Chiba, and H. Ohno, 
Appl. Phys. Lett. {\bf 79}, 2776 (2001).

\bibitem{pask}
J. E. Pask, L. H. Yang, C. Y. Fong, W. E. Pickett, and S. Dag, 
Phys. Rev. B {\bf 67}, 224420 (2003).
                                                                                
\bibitem{galanakis}
I. Galanakis and P. Mavropoulos, 
Phys. Rev. B {\bf 67}, 104417 (2003).

\bibitem{liu}
Wen-Hui Xie, Ya-Qiong Xu, and Bang-Gui Liu, D. G. Pettifor,
Phys. Rev. Lett. {\bf 91}, 037204 (2003).


\bibitem{PBE}
J. P. Perdew, K. Burke, and M. Ernzerhof, 
Phys. Rev. Lett. {\bf 77}, 3865 (1996).

\bibitem{fong1}
C. Y. Fong, M. C. Qian, L. H. Yang, J. E. Pask, and S. Dag, 
Appl. Phys. Lett. {\bf 84}, 239 (2004).

\bibitem{fong2}
C. Y. Fong and M. C. Qian, 
J. Phys.: Conden. Matter {\bf 16}, S5669 (2004).

\bibitem{fong3}
M. C. Qian, C. Y. Fong, W. E. Pickett, J. E. Pask, L. H. Yang, and S. Dag, 
Phys. Rev. B {\bf 71}, 12414 (2005).

\bibitem{fong4}
M. C. Qian, C. Y. Fong, W. E. Pickett, and Huai-Yu Wang, 
J. Appl. Phys. {\bf 95} 7459 (2004).

\bibitem{ssan}
S. Sanvito and N. A. Hill, Phys. Rev. Lett. {\bf 87}, 267202 (2001).
                                                                                
\bibitem{nanjing}
Hongming Weng, Jinming Dong, Phys. Rev. B {\bf 71}, 035201 (2005).

\bibitem{bolduc}
M. Bolduc, C. Awo-Affouda, A. Stollenwerk, 
M. B. Huang, F. G. Ramos, G.
Agnello, and V. P. LaBella, Phys. Rev. B {\bf 71}, 033302 (2005).

\bibitem{du}
F. M. Zhang, X. C. Liu, J. Gao, X. S. Wu, Y. W. Du, H. Zhu, J. Q. Xiao, 
and P. Chen, Appl. Phys. Lett. {\bf 85}, 786 (2004).

\bibitem{kresse}
G. Kresse and J. Hafner, J. Phys.: Conden. Matt.  {\bf 6}, 8245 (1994); 
G. Kresse and J. Furthmuller, Phys.Rev. B {\bf 54}, 11169 (1996).

\bibitem{van}
D. Vanderbilt, Phys. Rev. B {\bf 41}, R7892 (1990).

\bibitem{qian}
M. C. Qian {\it {\it et al.}} (unpublished).

\bibitem{monk}
H. J. Monkhorst and J. D. Pack, Phys. Rev. B {\bf 13}, 5188 (1976).

\bibitem{hor}
C. Hordequin, D. Ristoiu, L. Ranno, and J. Pierre, Eur. Phys. J. B {\bf 16}, 287 (2000).

\end{thebibliography}
\end{document}